\documentclass{sigchi-ext}
\usepackage[T1]{fontenc}
\usepackage{textcomp}
\usepackage[scaled=.92]{helvet} % for proper fonts
\usepackage{graphicx} % for EPS use the graphics package instead
\usepackage{balance}  % for useful for balancing the last columns
\usepackage{booktabs} % for pretty table rules
\usepackage{ccicons}  % for Creative Commons citation icons
\usepackage{ragged2e} % for tighter hyphenation
\usepackage[subtle]{savetrees}
\usepackage[utf8x]{inputenc}

\def\plaintitle{``So, tell me what users want, what they really, \emph{really} want!''} 
\def\emptyauthor{}
\def\plainkeywords{Preference elicitation; well-being; values in design; eudaimonic and hedonic UX.}

\title{``So, Tell Me What Users Want, What They Really, \emph{Really} Want!''}

\numberofauthors{4}
\author{%
  \alignauthor{%
    \textbf{Ulrik Lyngs}\\
      \affaddr{Dept of Computer Science}\\
      \affaddr{University of Oxford}\\
      \email{ulrik.lyngs@cs.ox.ac.uk} }
  \vfil \alignauthor{
    \textbf{Reuben Binns}\\
    \affaddr{Dept. of Computer Science}\\
    \affaddr{University of Oxford}\\
    \email{reuben.binns@cs.ox.ac.uk} }\\
  \vfil \alignauthor{
    \textbf{Max van Kleek}\\
    \affaddr{Dept. of Computer Science}\\
    \affaddr{University of Oxford}\\
    \email{max.van.kleek@cs.ox.ac.uk} }\\
  \vfil \alignauthor{
    \textbf{Nigel Shadbolt}\\
    \affaddr{Dept. of Computer Science}\\
    \affaddr{University of Oxford}\\
    \email{nigel.shadbolt@cs.ox.ac.uk} }
  }

\definecolor{linkColor}{RGB}{6,125,233}
\hypersetup{%
  pdftitle={\plaintitle},
  pdfauthor={\emptyauthor},
  pdfkeywords={\plainkeywords},
  bookmarksnumbered,
  pdfstartview={FitH},
  colorlinks,
  citecolor=black,
  filecolor=black,
  linkcolor=black,
  urlcolor=linkColor,
  breaklinks=true,
}

\begin{document}

%% For the camera ready, use the commands provided by the ACM in the Permission Release Form.
\CopyrightYear{2018} 
\setcopyright{acmcopyright} 
\conferenceinfo{CHI'18 Extended Abstracts,}{April 21--26, 2018, Montreal, QC, Canada \\
\copyright~2018 Association for Computing Machinery.}
\isbn{978-1-4503-5621-3/18/04}\acmPrice{$15.00}
\doi{https://doi.org/10.1145/3170427.3188397}
% Then override the default copyright message with the \acmcopyright command.
\copyrightinfo{\acmcopyright}

\maketitle

% Uncomment to disable hyphenation (not recommended)
% https://twitter.com/anjirokhan/status/546046683331973120
\RaggedRight{} 

% Do not change the page size or page settings.
\begin{abstract}
  Equating users' true needs and desires with behavioural measures of 'engagement' is problematic. However, good metrics of 'true preferences' are difficult to define, as cognitive biases make people's preferences change with context and exhibit inconsistencies over time. Yet, HCI research often glosses over the philosophical and theoretical depth of what it means to infer what users really want. In this paper, we present an alternative yet very real discussion of this issue, via a fictive dialogue between senior executives in a tech company aimed at helping people live the life they `really' want to live. How will the designers settle on a metric for their product to optimise?
\end{abstract}
%Nevertheless, designers and engineers routinely need to infer the true goals of their users, while 


  \category{H.5.m.}{Information Interfaces and Presentation
    (e.g. HCI)}{Miscellaneous}{}{}
%PAPER CAN BE MAX 10 PAGES INCLUDING REFERENCES
\section{Introduction}
  %From binge-inducing media recommender systems to politically-polarising newsfeed algorithms, tech companies are increasingly criticised for optimising for clicks rather than the best interests of their users \cite{Wu2016, Alter2017, Leslie2016, Cerf2011}. However, it is surprisingly difficult to define what inferring and optimising for users' 'best interests' might actually mean in practice \cite{Pommeranz2012,Zimmerman2009}. 
    The question of how to bridge the potential gap between what users~\emph{actually} do and what they~\emph{`really wanted'} to do has a relatively long history in human-computer interaction research. In the 1960's, Warren Teitelman's `Do What I Mean' (DWIM) philosophy argued that systems should not just execute whatever potentially erroneous instructions users put into a terminal ~\cite{Teitelman1966}. Instead, they should try to interpret users' true intentions and correct their errors (the implication being DWIM, Now What I Say (or Do)). In practice, Teitelman's error-correction systems were critiqued as merely reflecting what their~\emph{designer} would have meant (`do what Teitelman means') \cite{Steele1996}.
    
    The issue crops up in more fundamental ways in the domains of decision support and recommender systems, where the gap is not just between what the user typed and what they really intended, but between recorded interaction behaviour and what can be inferred about the user's wants and needs. On those rare occasions where HCI researchers in such fields venture into the moral minefield of defining 'what users really want', they often provide definitions which are intuitively reasonable yet dis-satisfyingly cursory; the related philosophical debate is then swiftly swept under the carpet.
    
    For instance, in an otherwise enlightening chapter, Jameson et al. offer the following on what decision-support systems should optimise for: ``... a `good outcome' [is] one that the chooser is (or would be) satisfied with in retrospect, after having acquired the most relevant knowledge and experience. Admittedly, this assumption is subject to debate...'' ~\cite[p.~35]{jameson2014choice}. 
    
	Similarly, Pommeranz et al. write in relation to methods for eliciting user preferences: ``More research is needed to design preference elicitation interfaces that elicit \emph{correct} preference information from the user'' \cite[p.~361]{Pommeranz2012}. Later, they consider what normative aspects might be required to determine such `correctness' of a preference and find answers to this question in short supply: ``There is much room for more explicit consideration of human preference construction also including values and affective aspects'' \cite[p.~365]{Pommeranz2012}.
  
Parallel questions are arising in recent AI research. For instance, techniques like `inverse reinforcement learning' (IRL) attempt to infer an underlying goal function from behavioural output \cite{Ng2000}. It is implied that such goal functions are equivalent to the `true desires' of the human from whom the machine is learning. Indeed, as Stuart Russel put it, a well-aligned AI `will watch all of us to learn more about what it is that we really want''.\footnote{Talk at TED2017 \url{https://www.ted.com/talks/stuart_russell_how_ai_might_make_us_better_people/transcript}} A common assumption in these techniques is that humans make optimal decisions, with deviations from optimality reflecting `random noise' in action selection \cite{Evans2016}. However, real human decision-making deviates systematically from optimality, because of cognitive biases like asymmetric perception of losses and gains, which make people sensitive to how identical outcomes are framed, or hyperbolic discounting of future rewards, which often make people inconsistent over time \cite{Tversky1974,Kahneman1991}. A rational model may wrongly assume that the true preference of, for example, a smoker who is trying (and failing) to quit is to smoke \cite{Evans2016}.

Indeed, some recent work in this space acknowledges the limitations of the pure behaviourist dream. Armstrong et al. argue that current IRL methods  are ``fundamentally and philosophically incapable of establishing a `reasonable' reward for the human'', which can only be overcome, they argue, by building in ``normative assumptions about the reward and/or planner'' ~\cite[p.~2]{Armstrong2018}. However, the authors stop short of articulating what those normative assumptions might be.

  An alternative approach is to devolve responsibility for those normative assumptions back to the user. 
  %Research on methods and interfaces for eliciting people's preferences has been ongoing for decades \cite{Peintner2008,Gajos2005}. 
  Numerous ways of explicitly eliciting users' preferences have been explored in the literature, including user ratings and example-critiquing in recommender systems \cite{Pu2009,Pommeranz2012} and absolute measurement and pairwise comparison in decision support systems \cite{Aloysius2006,Chen2004}. Moreover, pioneering work in this space explored ways to let users inspect and tweak a system's model of them \cite{Cook1994,Kay1997,Pu2009,Kay1995}. 

  However, explicit methods also run into problems. Foremost, the elicitation process itself greatly influences what users say they want. For example, users may prefer different options based on whether they are framed as losses or gains \cite{Pommeranz2012}, or depending on the moment in time in which they are asked. What point in time reflects what the user 'really' wants - the most recent, a weighted average over the last day / week / year, or something else? \cite{Kahneman2005,Redelmeier1996a}.
  
  Related to the question of timeframes is the tension between broad or narrow construals of the user's context. One of the foundational tenets of user-centred design, as articulated by Ritter et al., is to consider the user context more broadly ~\cite{ritter2014user}. That is, moving beyond the immediate, task-related issues pertaining to a specific product, where the user's goals can be more easily approximated, and instead view applications of technology as "the development of permanent support systems and not one-off products that are complete once implemented and deployed" ~\cite[p.~44]{ritter2014user}. In other words, the designer should consider longer-term effects of systems on people's lives, which in turn requires deeper insight in order to align systems with users' more general goals, values, and life situation.
  
  %The question of how to infer what users really want - as opposed to what might be inferred from what they actually say or do - is highly practical, as solutions are routinely implemented by system designers. And yet, as we hope to show in this paper, it is also a rabbit hole of theoretical and philosophical complexity.

The question of how to elicit 'correct' preferences therefore ends up connecting with the philosophical debate on the fundamental constituents of `what makes someone's life go best'. While it's possible that one's best life could be at odds with one's desires, it is often held that being able to satisfy `true desires' are at least partly constitutive of the good life. Most philosophers' answers tend to involve one or more of the following: pleasurable experiences or \textbf{`hedonism'} (e.g.~\cite{bentham1996collected}), where a good life is one full of pleasurable experiences; \textbf{desire-satisfaction} (e.g.~\cite{parfit2012makes}), where a good life is one in which one's desires are fulfilled; and \textbf{objective list} theories, where a good life is one in which certain objectively worthwhile things are experienced, achieved or engaged in (e.g. ~\cite{hurka1993perfectionism}). 
%According to both hedonism and desire-fulfillment, one's `best life' is at least partially determined by pleasure and desire; but most philosophers who subscribe to such accounts borrow from the objective list theory the notion of some `objective good'. 
Hedonists often concede that certain hedonic pleasures are more truly constitutive of a good life, and proponents of desire-fulfillment acknowledge that the fulfillment of misguided desires contributes little to one's best interests. As we hope to show, these philosophical debates underlie some of the conceptual difficulties encountered by the more practically-oriented research within the HCI / AI research cited above.

   To illustrate this, we now present an alternative yet very real take on the issue, via a fictional dialogue between senior executives in a global tech company whose product aims to help people live their best possible life. How can they settle on the metric their product should optimise so that they give their users what they \textit{really} want? In addition to the literature reviewed so far, the positions of our fictive designers are inspired by common views within classical economics and rational actor models \cite{McFadden1999}, behavioural economics \cite{Tversky1974,Kahneman1991}, and philosophical and psychological work on the 'good life' \cite{Seligman2005,Diener2002}, as well as common ideological stances. We acknowledge that this is only a portion of the relevant literature, and we have taken artistic license in our formulations of each stance. As such, we cannot claim our fictive characters to be fully and fairly representative of the possible range of positions. Furthermore, any resemblance to real persons is probably not accidental.
  
%TODO: wrap the character's base thoughts into a mini 'twitter profile' for each
%We base the opinions of our imaginary designers on commonly encountered opinions in the scholarly literature and tech communities. As will be clear throughout the acts, Randy Na represents the position that the best we can do for people is to maximise free choice so they themselves can choose what is best for them. This position has roots in classical economic models of rational actors and in libertarian variations of political philosophy (REFs). Harald Richter is sceptical of the idea that maximising choice is a good thing - even though people know what they want, they are often not able to act in their own best interests, because their environments tempt them in the wrong direction. This position has roots in behavioural economists’ and psychologists’ work on decision biases, and in the ‘nudge’ movement (REFs). Finally, Nichola Machian takes the position that people can’t know what is good for them - rather, we need guidance from those older and wiser or from philosophers or religions to know which things really make life worth living. This position has roots in philosophical work on the good life and scholars of religion (REFs).

\section{What Users Really Want: The Tale of Gamaface}
  \textbf{SCENE}: The Californian morning sun shines through the gleaming windows into a pristine meeting room at the headquarters of Gamaface, a global technology company. Gamaface is the industry leader in what they call 'algorithmic life services'. Their augmented reality platform - which combines decision support, persuasive computing, and ubiquitous personalised nudging - is used by 3 billion users 24 hours a day, 7 days a week.

  \textbf{CAST}:
      \begin{itemize}[topsep=0pt]
        \item Sunny Zuckerbezos, CEO of Gamaface \\
        @Walkin\_on\_sunshine | Mission: connecting people and data to make life worth living.
        \item Randy Na, Information Architect \\
        @Nozick\_SoSick | Libertarian | Autono-me, autono-you | "I hold it to be the inalienable right of anybody to go to hell in his own way."
        
            %POSITION: RADICAL LIBERTARIAN - PEOPLE KNOW WHAT IS GOOD FOR THEM: people are completely free - if we can get them to do something, it must be because they want to do it. If they don't like our product then they'll use something else
            %UL: Randy Na will argue that we should maximise choice so that people can do whatever they want. The outcome metric is probably purely behaviouristic, because people's choices reflect what they really want. If they do more stuff, then it must be because we are giving them more of what they want.
            %UL: position is very widely held in political debates and pushed by silicon valley; academically, this is the old school rationalist economist position

        \item Harald Richter, User Researcher \\
        @WinkWinkNudgeNudge | Pavlov's dog, striving to be Pavlov's bell. \#cognitivebias \#positivepsychology
        
        %POSITION: NUDGE THEORIST - PEOPLE KNOW WHAT IS GOOD FOR THEM, BUT NEED SCAFFOLDING; WE NEED TO HELP THEM TO HELP THEMSELVES
            % name is anagram of richard thaler
            %UL: Harald will argue that we should ask people to reflect on what they want and then help nudge them towards behaviour consistent with their personal preferences.
            %UL: position is held by behavioural economists and widely in psychology and HCI

        \item Nichola Machian, Lead Ethicist \\
        @eudaimonia\_for\_all | "Meaning arises when subjective attraction meets objective attractiveness"
          
          % POSITION: (NOT ONLY DO PEOPLE NOT KNOW WHAT IS GOOD FOR THEM, THEY ARE ALSO WILDLY INCONSISTENT OVER TIME - WE NEED TO SAVE THEM FROM THEMSELVES)
%           position: what's good for people is perfecting their human nature
          % name: perhaps find an anagram of Susan Blackmore, Sam Harris, Derek Parfit? Nichola Machian - a nod to Aristotle's Nichomachean Ethics
          %UL: position is probably held by objective list-philosophers

          %UL: It's not obvious where the social constructivist position fits in - this position would argue that humans have nothing that's innately good for them, but that all our needs are constructed by society. Maybe Randy Na could bring this up, because one can argue that the implication is that we need a very deep understanding of a person's culture and individual history to know what is good for them, and that we are therefore better off leaving people to their own devices?
          
    \end{itemize}

  \subsection{\textbf{ACT}: In search of a metric}
      \textbf{Sunny}: So, I've called you all here to pitch ideas for a new metric that will drive the direction of our services for the future. As you know, millions of people rely on our technology to guide their every waking (and sleeping) minute. The Gamaface mission has always been about our users ... giving them what they want, helping them to live better lives, the lives they really want to live. Across all our user's devices - laptops, smartphones, smartwatches, and smart glasses - our activity feed suggests what they should do next to live their ideal life. We help them find stuff that's 'relevant', see their most 'engaging' videos, and ping them 'helpful' info when they need it. 
      
      But what does any of that actually mean? How can we be sure that we are giving users what they really want? What we need, my friends, is a clear answer to this question; a new metric towards which all our services should be geared; a new optimisation metric for life. So come on, hit me with your ideas!

      \textbf{Randy}: I'm going to stop you right there, sir, if I may. What's wrong with our existing systems? We infer what users want from what they do and what other people like them do. If they spend every spare second watching cat videos, then our algorithms should give them more cat videos. If they keep watching them, that means our algorithms got it right. If they don't like them they will stop looking at them. Our algorithms will then show them less in the future ...

      \textbf{Harald}: Woah there. I totally disagree. People are slaves to simple reward functions inherited from our evolutionary past. We know how to hack these reward systems, so if we leave people to their own devices (no pun intended) they will simply do whatever our algorithms nudge them to do. That might be binge-watching cat videos and ordering takeout pizza. It probably won't be filling in their tax returns or exercise ...
      
      \textbf{Nichola}: But we could be nudging them to do those things instead! Even better, we could nudge them to do something truly worthwhile, like reading poetry, or contributing to science, or meditating on the miracle of their very existence!

      \textbf{Randy}: How patronising! As cosmopolitan, liberal, college-educated, Silicon Valley elites, who are you to decide what people should be 'nudged' towards? Sounds pretty paternalistic to me!
      
      \textbf{Sunny}: OK people, let's work together here. On the one hand, Harald and Nichola are on to something: our algorithmic life services shouldn't feed people's worst habits. But Randy also has a point; Gamaface must remain a neutral platform, with no political, ethical or aesthetic biases. All our algorithms are just based on pure mathematics and user behaviour - not on force-feeding them a specific notion of the good life.
      
      \textbf{Randy}: Exactly. The point of our service is to free people from the tyranny of the structures that control their lives - the government, social norms, the media - and let them do whatever~\emph{they} want to do. If 99\% of people choose to indulge what you call their `worst habits', that's their prerogative and we should help them. Equally, the 1\% who want to waste their time reading poetry are entitled to do so!
      
      \textbf{Nichola}: Are you really saying you believe that there is no objective difference between the aesthetic value of cat videos and Shakespeare?
      
      \textbf{Randy}: My opinion of the value of anything isn't the point here. The point is that none of us get to decide that for someone else!
      
      \textbf{Nichola}: Ah, but you'll agree that each individual may value different activities or pursuits as 'higher' and 'lower'? And if someone wants to pursue something they think is worthwhile - say, reading poetry - we should help them?
      
      \textbf{Randy}: Mhmmm ...
      
      \textbf{Nichola}: So if they wish they would read more poetry, but find themselves watching cat videos, we should stop the cat videos and replace them with poetry!
      
      \textbf{Randy}: Nope! People gotta live with the consequences of their failure to live up to their ideals!
      
      \textbf{Harald}: Aha! You're admitting that people might sometimes end up doing things they don't~\emph{really} want to do! In fact, this is a well documented phenomenon in behavioural science. Our system 1 - the fast, powerful, routine animal part of our brain - wants one thing, and system 2 - the slow, rational, reflective human part - wants another. Unfortunately, even though people identify with their system 2, system 1 usually gets its own way.
      
      \textbf{Randy}: Well, I don't know about that system 1, 2 mumbo-jumbo - speak for yourself pal, my systems are all fine, thank you very much. I guess maybe sometimes people are conflicted about what they want ... but even then, who are you to interpret which desire should prevail - which system represents what they~\emph{really} want?

  \textbf{Nichola}: We shouldn't decide between us in this room, but we could let the wisdom of science and philosophy decide! Psychologists have spent decades figuring out which activities and circumstances make people happy with their lives, and philosophers have pondered it for millennia. Shouldn't our systems help people live the way experts say make a life go well?
    
      \textbf{Randy}: That's absurd! The world is changing faster than ever before, and you suggest that some dead navel-gazing philosophers know what's best for people? Or maybe that we nudge everyone to get married if psychologists say married people are happier on average? What someone really wants is a function of their free will and their unique personality, and no-one besides themselves are in a position to know what's best for them ...
    
    \textbf{Nichola}: Are you seriously denying that things like close friends, meaningful work, good physical health, and feeling competent and valued in your community aren't universal constituents of a good life? It seems pretty obvious that people have fundamental - and universal - needs that must be satisfied for them to feel their lives are going well. We should help them get what will make them happy and fulfilled, even if they haven't themselves realised what the important things in life are ...

    \textbf{Harald}: But people also want to feel that they're in control of their lives - we can't dictate to them what they should want. Hmm... but what we \textit{could} do is help them reflect on what they want from life and then let them set the nudges they need from our products accordingly ...
    
    \textbf{Randy}: For God's sake... People don't want to live like saints! It's all very well to meditate on what your `ideal self' might want but sometimes people just want to indulge and we shouldn't make them feel bad about that! Besides, who wants to be forced to reflect like some Buddha on your life? Most people are lazy. %Surely we could just ask people now and then whether they are currently doing what they'd prefer to do, given the options they have available?
    
    %whether they feel like they're freely choosing to do what they're currently doing?
    %"they think they're doing what they really want to be doing." - changed as per max' critique
    
    \textbf{Sunny}: OK, there's a lot of stuff here. Let's come back to my original question to you all. Given everything you've said, what is the metric you think we should optimise, and how do we do it? %It sounds convincing that there's some basic conditions for the good life that hold true across all users, as Nichola says. But it's also true that people don't want to be told what to do, especially not by a Silicon Valley elite. So whatever we try to optimise in people's lives, it should obviously be supported by users themselves. Perhaps we can discover this by encouraging user's to reflect, as Harald suggests, or by inferring their higher wants - and the mix they want between different pleasures - from their behaviour, as Randy says.
    
    \subsubsection{Metric \#1: Engagement with preferred options} %should this be 'autonomy' or 'free choice' or 'free will'
    \textbf{Randy}: Fundamentally, we must believe that our users can choose for themselves what the good life is - anything else is frankly disrespectful. So our metric should be rooted in behaviour: If people engage more with the options we give them, it must, all else being equal, be because they find them valuable. `But what if they're addicted', you say? We correct for that by asking them now and then whether they are currently doing what they most want to be doing given the their options. Surely, if they're addicted, their answer will be 'no'. So we use two metrics: what they \textbf{actually do} and \textbf{what they in the moment say they most want to do}. When we optimise both, then we give them what they really want.
    %prefer what they currently want. what they're currently doing is out of their own free choice - 

  So basically, we're almost there. Our apps and augmented reality systems already put the options users engage with the most in front of their eyes and fingertips. %The videos they watch the most, the people they communicate with for the longest and most frequently, the games they’re most engaged with. We only need a minor, optional addition to this system, which is to sample how they feel in various moments, so that we know that they’re engaging with things that makes them their happiest rather than just things that they’re addicted to. 
    But we build a little extra that we could brand \textit{Gamaface Autonomy Sense\textsuperscript{TM}}: a simple overlay that occasionally asks our users, on a scale from 1-10, to which degree they are currently doing what they most want to be doing given their options. We get responses across a user's different activities, add some random noise to the activities we recommend, et voilà! We learn the activity schedule that optimises the user's engagement with preferred options.
    
    \subsubsection{Metric \#2: Regret when reflecting on past activity}
    %HARALD: IDEAL SELF REINFORCEMENT LEARNING / PURE SELF-CONTROL, EXPLICIT RULES 
          %RL where user puts themself into 'ideal self' mode, then does reinforcement learning.  provide items they’ve consumed / read / watched and rate them according to ‘actual self’ and ‘ideal self’ alignment - testing different ways of eliciting the ideal self. then see if models can predict each type of rating.
          %Pure self-control, explicit rules - let people write their own hand-coded rules like 'only show me 10 cat videos per week', 'order web pages by pagerank during the week, but by serendipity at the weekend'.
      %\textbf{Harald}: No, no, no, we can do much better than this. We need our systems to give space where our users explicitly reflect back on their past choices and indicate which they were happiest with. 
            %\textbf{Harald}: But life isn't just about having enjoyable experiences. It's about doing things you find meaningful too. If you're constantly regretting your past behaviour, that's probably because it's not meaningful enough!
     
   \textbf{Harald}: But people often regret their past behaviour, even if they thought they wanted it in the moment! Really, what we should do is help them get as close to how they wish to be when they take the time to reflect. Therefore, our metric should be \textbf{amount of regret when the user carefully considers his past activity}. When we minimise this regret, then we are giving users what they really want.
    
  So we build a system we brand \textit{Gamaface Deep Ideal Self Learning\textsuperscript{TM}} - this is guaranteed to play well with our audience! It lets the user review a timeline of her past behaviour. For each activity, or activity class, she rates on a scale from 1-10 how closely what she did matches what she'd like her ideal self to do, and/or how often she'd like her ideal self to do it. This will allow our algorithms to learn how to present activities and nudge her so that her actual behaviour gets closest to her ideal self.

    %\textbf{Randy}: So if someone is watching cat videos, and enjoying it in the moment, you want our service to interrupt them and say: `Are you sure you want to be doing this? Last time, you regretted wasting time this way! Here's some poetry to read instead'. How paternalistic, how infantilising! Respecting users means treating them like adults, whether they regret their actions later or not. You are assuming that our `real' identity is revealed only when we reflect on our past behaviour, but isn't what we choose to do in the moment more reflective of who we `really' are? If you spend your time doing things that you enjoy in the moment, maybe you should accept you're someone who just loves cat videos and there's nothing wrong with that! Stop regretting, just enjoy yourself!
    \textbf{Randy}: This makes my skin crawl. First, it doesn't matter what people \emph{would} have chosen under `ideal' reflective circumstances, whatever they might be (let me guess; sitting cross-legged like a Zen master while soothing dolphin noises play in the background?). Second, it's disrespectful to override someone's present desires in order to serve some inferred `ideal' desires. Finally, the random, irrational, and downright~\emph{messy} nature of the human condition is what makes life worth living.
    
    \textbf{Nichola}: Well, that explains all the garbage piled up around your desk. %I think you're both right and wrong. Harald is right that a good life contains both enjoyment ~\emph{and} meaning. Randy is right that we shouldn't place too much weight on what people regret about their past behaviour. 
    Harald, I think Randy is right that we shouldn't place too much weight on what people regret about their past behaviour. Regret is an imperfect guide; the grass is always greener on the other side, and maybe that goes for deathbed lamentations too. Who's to say that you wouldn't have regretted an alternative life course even more? And if you don't regret your actual life choices, maybe that's because you don't appreciate what could have been! Who knows which of all your possible future selves would have the `best life'?
    %We can't simulate an individual's multiple possible life trajectories and pick the best one. UL: I've taken this sentence out because if you try to crowdsource the problem and make interpersonal comparisons, then it seems to me that you're effectively trying to simulate and individual's multiple possible life trajectories, aren't you?
%     So we can't assume that any individual has the wisdom, self-knowledge, or time to know what is really good for them.
    
    \textbf{Sunny}: So are you saying it's impossible?
    
    \textbf{Nichola}: Well, not entirely. I think the solution is to draw on wisdom outside of the individual - putting the burden of figuring out what the good life is on the shoulders of each user is frankly setting them up for failure. 
    
    \subsubsection{Metric \#3.1: Similar, wiser users' engagement and regret}
    %NICHOLA: System informed by the position that people have little idea of what's actually good for them. This system takes its input from the accumulated wisdom of its oldest users as to what would have been the best way for them to lead their lives, based on a weighted contribution of ongoing experience sampling throughout their lives as well as their retrospective judgments. The system uses this information to set default nudge values for its youngest users - 'based on your life situation, most similar users found that these choices lead to the most optimal life'. Alternatively, this system could ignore user judgments and experiences altogether, and take its settings from expert philosophers or, alternatively, a religious or other ideological framework
    
    \textbf{Nichola}: The first option is learn from our users' aggregated wisdom, and use comparisons across different people's real lives as they actually lived them to infer the necessary conditions for a happy and meaningful life. So we can use Randy's and Harald's suggested systems to collect data, but the goal is to create metrics that refer to \textbf{the collective elicited preferences of similar users with more life experience} instead of just whether the user himself thinks something is good for him. We are most likely to give users what they really want when we, based on the accumulated life wisdom of others, optimise expected engagement with preferred options and minimise expected subsequent regret.
    This system we call \textit{Gamaface Wisdom of Age\textsuperscript{TM}} - a collaborative filtering approach to the good life, in which each user's contribution is weighted by their experience. We set default choices and nudges of our youngest users based on the wisdom of more experienced users. 
    %what would've made our older users most engaged, autonomous, and satisfied after the fact, measured as Randy and Harald suggest.
    %Our metric is the amount of \textbf{engagement with momentarily preferred options, and subsequent regret that other, but more experienced, users went through in similar situations}. We are most likely to give users what they really want when we optimise expected engagement and experienced autonomy, and minimise expected subsequent regret, based on the accumulated life wisdom of others. 

    \subsubsection{Metric \#3.2: Alignment with guru-guided good life}
    The other option is to use some expert guidance on the life, or what we might call the 'wisdom of gurus': then our metric is \textbf{alignment between a user's circumstance and what the guru system says leads to an enjoyable/meaningful life}. 
    
    Certain conditions have been found by psychological research to make everyone miserable, so some of the guru reference values should be the same for everyone. On the other end of the spectrum, there might be many different ways to make a life go really, really well, so here we could let users choose which guru system they prefer, like their favourite philosopher, religion, or other ideology. We give users what they really want when we minimise the conditions in their lives that reliably make people miserable (e.g. loneliness) and optimise the conditions their guru system of choice says makes for an ideal life.
  
  This system we call \textit{Gamaface Guru-Guided Good Life (G4L)\textsuperscript{TM}}. We involve different experts directly in creating the values and nudges. We could also offer users an exclusive option of having our system trained on a text corpus from their philosophy of choice. Imagine that users can bring their favourite Buddhist texts--or your Ayn Rand novels, Randy?--and our machine learning algorithms will learn its values and calibrate the user's system accordingly?
    
    %We might need to consider boundaries, so that users actually stick with a value framework rather than just change it when they feel the urge to indulge something their current framework says they shouldn't... 
    %Perhaps this is really just an extension of raising a child, so maybe parents should have the say over the guiding framework for their children until the child reaches the age of consent.
    
    \textbf{Sunny}: Well, I'm going to have fun explaining all this to our shareholders... Time for lunch!

\section{Discussion}
  Given the problems with equating users' true needs and desires with simple behavioural measures of `engagement', alternative metrics are needed. In this paper, we have discussed this issue in the context of a dialogue between designers in a fictive tech company searching for a metric with which to measure whether they are giving users what they `really' want.

Despite the often glib treatment of these questions in HCI research, as seen in the examples in the introduction, there are some notable exceptions. The general critique of the behaviourist tendency in recommender systems is well-articulated by Ekstrand \& Willemsen~\cite{Ekstrand2016}. They suggest one promising corrective, which allows users to choose between different algorithms underlying their recommender systems \cite{Ekstrand2015}. In allowing some direct control of the inference process, this is reminiscent of earlier work which aimed to create `scrutable' user models which are transparent and configurable ~\cite{Kay1995,Kay1997}.

Harald's call for systems which help put users in a position from which they can reflect on themselves and their desires, has some precedent in Slovak et al.'s proposal to design for `reflective practicum'; a state in which someone can engage in transformative revision of their outlook or behaviour ~\cite{slovak2017reflective}. In addition, we suggest that empirical findings from positive psychology and reflections from the philosophy of what makes life go well will prove important as a source of more opinionated takes on how we might design to support people's `better selves' and draw on life wisdom that does not originate with the designer or individual user \cite{Seligman2005,Diener2002,Buss2000}.

Nichola's call to consider the~\emph{meaningfulness} of user experiences is something explored by Mekler and Hornbæk, who argue that UX designers should consider~\emph{eudaimonic} (`living the good and virtuous life') as well as~\emph{hedonic} (`pleasurable') experiences~\cite{Mekler2016}. Similarly, Zimmerman proposes `designing for the self', such that products can help their users become the people they desire to be~\cite{Zimmerman2009}.

For those sympathetic to Randy's position, the notion of an external force guiding users' search for the good life may detract from the sense in which an individual ought to be responsible for their own life journey. However, behavioural economists have found that the crucial question for whether people are in favour of `nudging' --in the sense of setting up the choice environments to support particular kinds of behaviour-- is whether or not they agree with the vision of the good life behind those nudges \cite{hagman2015public}. We can easily imagine a future in which users expect the right to choose which normative assumptions should be embodied in the algorithms feeding their recommender systems.

A central goal of this paper has been to show that whereas it is easy to criticise tech companies for equating users' true preferences with simple metrics of engagement, it is not obvious what good alternative metrics look like. Every metric implicitly embodies particular assumptions about human nature - and ultimately about the good life - with few decisive arguments for favoring one as the 'best'. It is a fallacy, however, to conclude that the question is therefore not worth bothering with and/or that all metrics are equally valid. Some ways of solving the problem are clearly worse - such as simply equating true preference with number of clicks or time spent using a service - even if there is no way to tell which of the better alternatives might be the global optimum. We believe that allowing users the to choose between different ways for a system to infer their preferences - for example, implementations of Randy's, Harald's, and Nichola's positions - would be a big step forward. To reach such a future, we need to properly engage with the issue and explore, build, evaluate, and discuss what good alternatives look like.

Some readers may still feel that this discussion is akin to moral philosophers' "trolley problem" - important in principle, but not terribly relevant in practice~\cite{Bauman2014}. However, even though recommender and decision-support systems typically used today are limited in scope to specific task and interest contexts, more sophisticated systems could begin to model their longer term effects on user's lives, and user's preferences towards such influence. In fact, such a future might rapidly be approaching. Addressing recent criticisms of Facebook, CEO Mark Zuckerberg recently announced that his personal challenge for 2018 included "making sure that time spent on Facebook is time well spent" \footnote{Facebook post on 4th January 2018 \url{https://www.facebook.com/zuck?hc_ref=ARQqfRj278TDWekby2TLyI0A0meA4\\-4PxqohaalwAfzCeAsMaft16fKBkDYiHEg4cQk&fref=nf}}. Following up a week later, Zuckerberg elaborated that his team felt "a responsibility to make sure our services aren't just fun to use, but also good for people's well-being" %. So we've studied this trend carefully by looking at the academic research and doing our own research with leading experts at universities"
and that he would be "changing the goal [of] our product teams from focusing on helping you find relevant content to helping you have more meaningful social interactions"\footnote{Facebook post on 11th January 2018 \url{https://www.facebook.com/zuck?hc_ref=ARTYPggwbi_dcIl5f8M8r1dGZYZGhpWmAXwj_9C6g6mmSCSxA0dpxqWqEaPojN1IWD0&fref=nf}}. Time will tell which, if any, of the paths suggested by Randy, Harald, and Nichola will be taken by world-leading social media platforms.

\newcommand{\showDOI}[1]{\unskip} %hide DOI's
\bibliographystyle{SIGCHI-Reference-Format}
\bibliography{what-users-really-want}

\end{document}